\documentclass{article}

\usepackage{arxiv}
\usepackage[utf8]{inputenc} 
\usepackage[T1]{fontenc}    
\usepackage{hyperref}       
\usepackage{url}            
\usepackage{booktabs}       
\usepackage{amsfonts}       
\usepackage{nicefrac}       
\usepackage{microtype}      
\usepackage{cleveref}       
\usepackage{graphicx}
\usepackage[square, numbers]{natbib}
\usepackage{doi}
\usepackage{subcaption}
\usepackage[ruled,linesnumbered]{algorithm2e}
\usepackage{multirow}
	
\title{Face: Fast, Accurate and Context-Aware Audio Annotation and Classification}

\author{\href{https://orcid.org/0000-0003-2237-1374}{\includegraphics[scale=0.06]
	{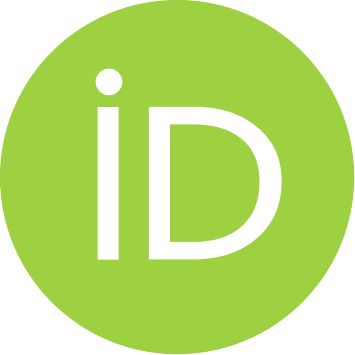}\hspace{1mm}M. Mehrdad Morsali}\\	
	Department of Electrical Engineering\\
	Sharif University of Technology\\
	\texttt{morsali@sharif.edu} \\
\And
\href{https://orcid.org/0000-0002-9852-5088}{\includegraphics[scale=0.06]
	{orcid.pdf}\hspace{1mm}Hoda Mohammadzade}\\	
	Department of Electrical Engineering\\
	Sharif University of Technology\\
	\texttt{hoda@sharif.edu} \\	
\And
\href{https://orcid.org/0000-0002-7715-8004}{\includegraphics[scale=0.06]
	{orcid.pdf}\hspace{1mm}Saeed Bagheri Shouraki}\\	
	Department of Electrical Engineering\\
	Sharif University of Technology\\
	\texttt{bagheri-s@sharif.edu} 
}

\renewcommand{\shorttitle}{Face: Fast, Accurate and Context-Aware Audio Annotation and Classification}

\hypersetup{
pdftitle={Face: Fast, Accurate and Context-Aware Audio Annotation and Classification},
pdfsubject={Audio Annotation and Classificaation},
pdfauthor={M. Mehrdad Morsali},
pdfkeywords={Active Learning, Sound Classification, Semi-supervised Audio Classification},
}

\begin{document}
\maketitle

\begin{abstract}
	
\end{abstract}
This paper presents a context-aware framework for feature selection and classification procedures to realize a fast and accurate audio event annotation and classification. The context-aware design starts with exploring feature extraction techniques to find an appropriate combination to select a set resulting in remarkable classification accuracy with minimal computational effort. The exploration for feature selection also embraces an investigation of audio Tempo representation, an advantageous feature extraction method missed by previous works in the environmental audio classification research scope. The proposed annotation method considers outlier, inlier, and hard-to-predict data samples to realize context-aware Active Learning, leading to the average accuracy of $90\%$ when only $15\%$ of data possess initial annotation. Our proposed algorithm for sound classification obtained average prediction accuracy of $98.05\%$  on the UrbanSound8K dataset. The notebooks containing our source codes and implementation results are available at \href{https://github.com/gitmehrdad/FACE}{https://github.com/gitmehrdad/FACE.}


\section{Introduction}
In recent years, sound event classification has gained attention due to its application in audio analysis, music information retrieval, noise monitoring, animal call classification, and speech enhancement \citep{Gururani, Wang, Shuyang}. The development in the mentioned fields requires annotated recordings \citep{Shuyang} and larger datasets instead of the smaller ones \citep{Fonseca}. Since labeling audio is time-consuming and expensive, establishing a large-scale labeled audio dataset is arduous \citep{Lee}. So, it seems necessary to develop reliable methods to label the datasets. There are several methods of determining the labels of unlabeled data. Some papers, including \citep{Gururani, Fonseca, Lee, Takagi, Chan, Benito} pursue Semi-supervised Learning, and some works, including \citep{Wang, Shuyang, Dilek, Malte, Qin, Ji} seek to utilize Active Learning. Regarding the superior annotation accuracy of Active Learning-based approaches, this paper also focuses on Active Learning to determine the labels of audio samples. In certainty-based Active learning, a small portion of data with deterministic labels trains a classifier at first. Then, the classifier labels the unlabeled data, and audio samples with the highest classification certainties join the training set. The cycle of classifier training, classification, and updating the training set continues until the complete annotation of all audio samples. 

Solutions proposed for the problem of audio event classification examine several methods. In \citep{Gururani, Fonseca, Lee, Chan, Benito}, they suggested utilizing deep multi-layer neural networks to classify the data. The main issue challenging some Deep Learning solutions is the low prediction accuracy, in the case of the model overfitting, due to the scarcity of training data. In the audio classification task, where usually a propitious amount of training samples are available, we won't face the model overfitting; however, in the audio annotation task, where the labels are available only for a small portion of data, the overfitting problem makes the deep multi-layer neural networks less attractive. In \citep{Wang, Shuyang, Takagi}, they have employed classic machine learning classifiers to classify the data. These methods are potentially less susceptible to overfitting since they have fewer unknown parameters that should get determined during the training. Nevertheless, the classic machine learning approaches suffer from the fixed input size issue. 

Regarding the considerable achievements of Context-aware Machine Learning \citep{Zeng}, this paper proposes some context-aware solutions for the addressed problems. A context-aware design considers the current situation for decision-making. As a case in point, based on the system's status, a context-aware Machine Learning system might dynamically renew the model fine-tuning and hyperparameter adjustment. The first research contribution of this paper includes a heuristic approach for context-aware feature selection to provide summarized fixed-size input vectors to gain a fast, accurate, and overfitting-resistant classification model. The investigation of the audio signal's Cyclic Tempogram feature would be another contribution of this work in the audio signal classification task. The proposal of a context-aware Active Learning-based method for audio signal annotation is the third research contribution of this paper. The fourth research contribution is the context-aware classifier design to obtain optimum classification results. Section \ref{method} describes the mentioned contributions in data preparation and feature extraction, data annotation, and audio classification. Section \ref{results} discusses the implementation results for the proposed methods. Section \ref{conclusion} concludes the paper.

\section{The proposed Method}
\label{method}

\subsection{Data Preparation and Feature Extraction}
\label{ablation}
Assume a dataset of audio signals with a duration of four seconds under the sampling rate of $22.05KHz$. Since each audio signal consists of $4\times 22050$ samples, the input vector of a classifier operating on that dataset should contain $88200$ variables. These variables might distinguish audio signals from each other; however, processing those lengthy input vectors requires large amounts of memory and expensive computational resources. Besides, as the classifier's input vector's length grows, the trained model tends to be more susceptible to overfitting \citep{Elements}. So, instead of regarding the time contents of an audio signal as the classifier's input vector, some distinctive features extracted from the signal's time contents would form the input vector. A context-aware feature extraction provides good classification accuracy and low model overfitting possibility. 

Some previous works, including \citep{Song, Theodoros, Su, Li}, discussed the various feature extraction methods applicable to the sound event classification task. Following a context-aware approach, we evaluate different feature extraction methods to find the most effective selection among that methods. Since this paper desires fast classification, we overlooked the evaluation of some methods with remarkable classification accuracy gain, like the Scattering Transform, because of their low feature extraction speed. The setup for the ablation study on the features contains a dataset, a feature extraction method, and a classifier. 

The employed dataset is UrbanSound8K (US8K) presented in \citep{Salamon}, which consists of $10$ classes of urban environment's audio events, including air conditioner, car horn, children playing, dog bark, drilling, engine idling, gunshot, jackhammer, siren, and street music. Each class consists of various counts of audio samples, and different-class data are arduous to distinguish for at least half the data classes. There are $8732$ audio samples with different sampling rates and a maximum duration of $4$ seconds. In this paper, the first step of data preparation is resampling the audio signals with the $22.05KHz$ sampling rate to obtain a uniform dataset. Data compression often ensues signal resampling since the sampling rate of audio signals usually exceeds $22.05KHz$. Since the US8K dataset does not specify a validation set, we regard $10\%$ of the training samples as the validation samples. All context-aware decisions in this paper are made based on the mentioned validation set. The examined features contain local autocorrelation of the onset strength envelope (Tempogram), Chromagram, Mel-scaled Spectrogram, Mel-frequency cepstral coefficients (MFCCs), Spectral Contrast, Spectral flatness, Spectral bandwidth, Spectral centroid, Roll-off frequency, RMS value for each frame of audio samples, tonal centroid features (Tonnetz), and zero-crossing rate of an audio time series. Regarding the lack of investigation history in the environmental sound classification papers, this paper studies the Tempogram, proposed in \citep{Grosche}, for the first time. 

Most of the mentioned feature extraction methods operate based on windowing an audio signal, so the extracted features are referred to as static since they represent the information of the signal's specific windows. The delta and delta-delta coefficients of the audio signal feature, which respectively indicate the first-order and second-order derivatives of the features, are referred to as dynamic features. The delta features show the speech rate, and the delta-delta features indicate the speech's acceleration \citep{Features}. Some feature extraction methods generate a Spectrogram, a visual representation of the audio signal's spectrum of frequencies. Regarding the Spectrogram as an image, we can exert some feature extraction properties from the Computer Vision scope, so we apply a Gaussian filter to the feature vectors before taking derivatives. In \citep{Elements}, feature selection helps to overcome the overfitting problem of the model by reducing the input vector's dimensions and to obliviate the fixed input size issue mentioned in the previous section. The input vector's length depends on the number of windows used during the feature extraction. When the windows' count gets too high, the information represented by consecutive windows won't differ substantially, and when the windows' count gets too low, the amount of the extracted features won't suffice to distinguish different audio samples easily. To reduce the data dimensions, the mean and the variance of the extracted features per window would replace the original data. A similar replacement happens for the delta and delta-delta vectors. So, the final feature vector embraces six vectors of floating-point numbers, containing the mean and the variance values for the original feature and first-order and second-order derivatives of it. In this paper, we used the feature extraction tools available by Librosa, a Python package, where most arguments of the feature extraction function are left untouched with their default values. Nevertheless, some function arguments, including the window length and the number of samples per window, got adjusted using the validation data. Before the experiments, we transformed features using Quantiles information into the standard vectors where the value of all elements lies in the range of (-1,1).

A context-aware choice for the classifier would be the one holding the best capability to distinguish the data. There is a dilemma in feature and classifier selection: a classifier might hide the distinction some features provide, while a feature set might hide the resolution some classifiers provide. So, initially, we explore various extracted features using a specific classifier chosen based on our speculation of the data distribution. After feature selection, we try to find the best classifier to distinguish the data. 

As the t-SNE plot for the UrbanSound8K dataset samples represented in \citep{Sun} shows, the data distribution in the features space is heterogeneous and sparse. In a heterogeneous data distribution, the neighborhood of each data sample embraces some data samples with different class labels. Sparse data have a distribution in which the same-class data samples don't shape dense areas. Assuming that the feature selection in our work also results in heterogeneous and sparse data distribution, we search for a classifier to separate the data classes. Classifiers drawing a linear decision boundary fail to separate sparse and heterogeneous data, and classifiers utilizing non-linear kernels require expensive computational effort. So, a decision tree classifier with an appropriate depth seems profitable. The high potential of model overfitting utilizing the decision tree classifier necessitates using an ensemble of tree classifiers. Despite the improvements, the model overfitting still overshadows the accuracy of the mentioned classifier, so the XGBoost classifier proposes a solution by moving the decision boundaries utilizing regularization \citep{Elements}. Inspired by the One-vs-All approach that avails the application of binary classifiers for multi-class classification, we propose to employ the One-vs-All XGBoost classifier. In fact, instead of a single multi-class XGBoost classifier, multiple binary XGBoost classifiers draw the decision boundaries, where each classifier tries to separate the members of a specific class from all other data. Regarding the light duty of a binary XGBoost compared to a multi-class XGBoost, the computational overhead of the proposed method would not be an issue.

\begin{figure}
	\centering	
	\includegraphics[width=\linewidth]{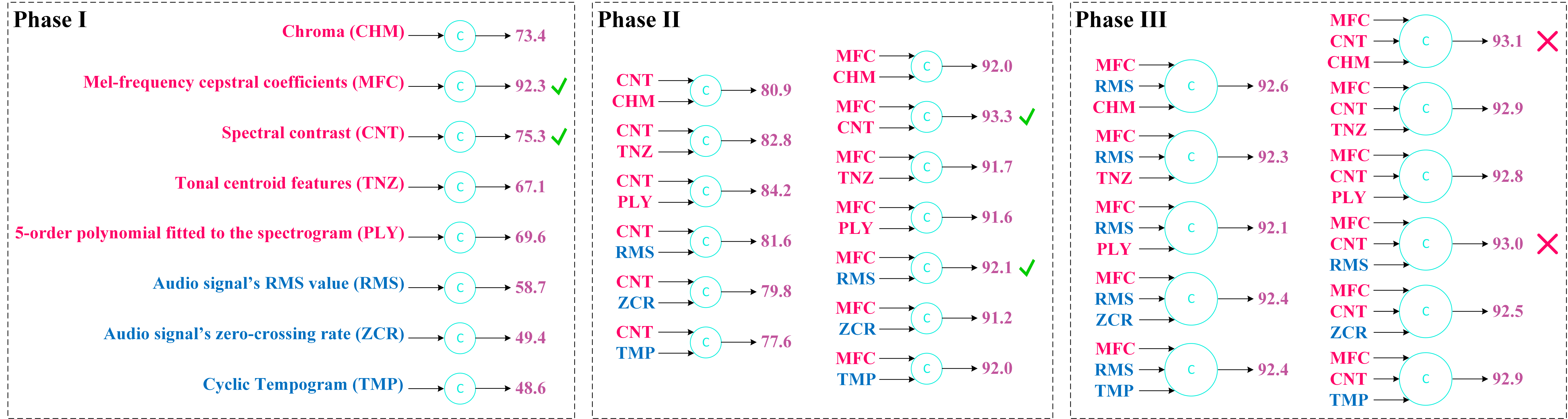}  		
	\caption{Feature selection resutls based on the proposed heuristic approach}
	\label{fig: fer}
\end{figure}
 
Figure \ref{fig: fer} depicts the feature selection results. To avoid the examination of all features permutations, we followed a heuristic approach, which consists of adding one new feature to each feature set at each phase, reporting the validation sample's classification accuracy, moving the top-two feature sets to the next study phase, and dropping further investigations when the classification accuracy stopped enhancement. In Figure \ref{fig: fer}, the results of Phase III don't show improvement compared to Phase II; therefore, we pick the MFC+CNT combination from Phase II as our final selection. Figure \ref{fig: fer} depicts methods' abbreviations in Phase II and Phase III of the experiment and shows the time-domain features in blue and the frequency-domain features in red. Figure \ref{fig: fer} doesn't represent the Mel-scaled Spectrogram feature extraction results since MFCC features are not only convertible to the Mel-scaled Spectrogram features but also capable of providing more classification accuracy. Some methods like Spectral flatness, Spectral bandwidth, Spectral centroid, and Roll-off frequency got excluded from Figure \ref{fig: fer} because of their poor classification accuracy results. Please note that the reported accuracies belong to a simple classification procedure where no Active Learning or other accuracy optimization techniques are applied. As seen in Figure \ref{fig: fer}, the remarkable accuracy the MFCC feature provides alone clarifies the reason behind the widespread use of that feature in most previous works. The feature vector of the selected combination consists of $810$ floating-point numbers, substantially reducing the computational resources required for the annotation and the classification tasks because it is sufficient to check $810$ features instead of $88200$ features. 

\begin{figure}
	\centering	
	\includegraphics[width=.5\linewidth]{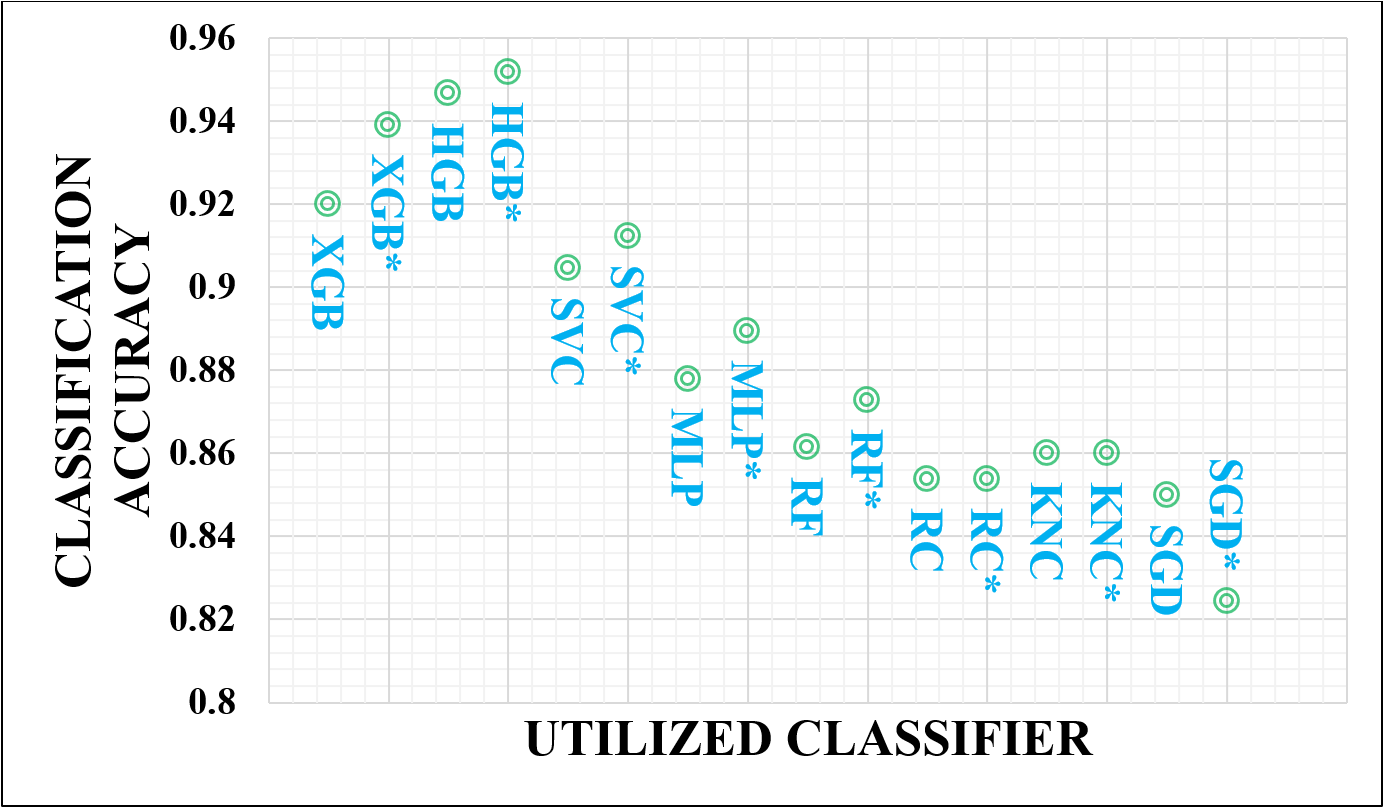}  		
	\caption{The classification accuracy vs. the utilized classifier}
	\label{fig: cas}
\end{figure}

Utilizing various classifiers, Figure \ref{fig: cas} shows the classification results where no Active Learning or other accuracy optimization techniques are applied. The experiment focuses on finding the best classifier for the annotation task, so it explores only the classic machine-learning methods. In Figure \ref{fig: cas}, XGB, HGB, SVC, MLP, RF, RC, KNC, and SGD, respectively, denote the XGBoost, the Histogram-based Gradient Boosting Classifier, the Support Vector Machine, the Multi-layer Perceptron, the Random Forest, the Ridge Classifier, the K-nearest-neighbors Classifier, and the Regularized Linear SVM with Stochastic Gradient Descent learning. In Figure \ref{fig: cas}, the asterisk sign after the classifier's name indicates using the classifier in the One-vs-All scheme. Each classifier uses our defined validation set as the test set and the rest of the samples in the US8K dataset as the training set. Figure \ref{fig: cas} shows the results in which classifiers succeed in obtaining at least $80\%$ classification accuracy, so it does not contain classifiers like Decision Tree, Gaussian Naive Bayes, AdaBoost, and QDA, which fail to reach $80\%$ classification accuracy. The results of Figure \ref{fig: cas} corroborate our speculation of the data distribution and the context-aware classifier selection to separate the classes. As shown in Figure \ref{fig: cas}, the HGB classifier outperforms the XGB classifier in terms of classification accuracy, but the XGB classifier operates faster. So, for the annotation tasks, we will use a One-vs-All XGBoost classifier, a context-aware choice that provides satisfying classification accuracy on the validation set. 

\subsection{Data Annotation}

 \begin{algorithm}
	\caption{The proposed method for audio event annotation}\label{alg: label}
	\KwData{$dataset$: A dataset with unlabeled data}
	\KwResult{$Label$: The predicted labels for the dataset}
	$Label \gets null$\;
	$L \gets dataset.SamplesCount$\;
	$LocalOutlierFactor(neighbors=1).Fit(dataset)$\;
	$I \gets LocalOutlierFactor.Inlier(size=L/40)$\;
	$R \gets RandomSelection(data=dataset, size=3L/40)$\;
	$Q \gets ActiveLearner.query(data=dataset, size=L/20)$\;
	$T \gets Concatenate(I, R, Q)$\;
	$Label[T] \gets AskLabel(T)$\;
	\For{$i\gets1$ \KwTo $4$ }{
		$Labeller.fit(dataset[T], P[T])$\;
		$Prediction \gets Labeller.Predict(dataset)$\;
		$Score \gets Labeller.PredictionScore(dataset)$\;
		\For{$j\gets1$ \KwTo $L$ }{
			$Outlier \gets LocalOutlierFactor.IsOutlier(dataset[j])$\;
			\If{$Label[j]=null$ and $Outlier=False$ and $Score[j]>0.7$}{
				$Label[j] \gets Prediction[j]$\;
				$T.Append(j)$\;
			}
		}
	}
	$Labeller.fit(dataset[T], P[T])$\;
	$Prediction \gets Labeller.Predict(dataset)$\;
	\For{$j\gets1$ \KwTo $L$ }{
		\If{$Label[j]=null$}{
			$Label[j] \gets Prediction[j]$\;
		}
	}
\end{algorithm}

For the annotation task, conventionally, it is assumed that a human annotator provides the labels for a small proportion of the data. Such an assumption leads us to avoid using deep neural network classifiers to annotate new samples because the low number of training samples and the high number of model parameters expose the model to overfitting, which in turn causes the prediction accuracy to drop. So, based on the results observed in Section \ref{ablation}, we will annotate the data using a One-vs-All XGBoost classifier, a context-aware choice that provides good classification accuracy. Another manifestation of a context-aware framework, realized in our proposed annotation method, is specifying the data the human labeler should annotate. Algorithm \ref{alg: label} shows our proposed annotation method. The Local Outlier Factor (LOF) method, proposed in \citep{Breunig}, finds the inlier samples having the least Euclidean distances from their nearest neighbor. The Active Learner, a Python tool from the modAL package, detects hard-to-predict data that are samples with minimum prediction certainty. According to what is shown from line $1$ to line $8$ of Algorithm \ref{alg: label}, a human annotator determines the labels for inlier samples, hard-to-predict samples, and some random samples, uniformly chosen from every different class. The maximum number of selected inlier, hard-to-predict, and random samples are respectively limited to $2.5\%$, $5\%$, and $7.5\%$ of the total sample count in the datatset. The philosophy behind choosing inlier samples for initial annotation is that there is a good chance that the neighborhood of the inlier samples embraces the same class samples. The reason behind adding some random data samples to the inlier samples is the possible imbalance in the count of data samples of different classes in the inlier samples. Another reason for adding random samples is to find different-class data samples adjacent to the inlier samples. 

In Algorithm \ref{alg: label}, line $9$ to line $27$ indicate the proposed context-aware Active Learning method. The method consists of five stages of data labeling where the predicted samples with reliable prediction accuracy also help determine the labels. In this work, a reliable prediction possesses at least $70\%$ prediction accuracy score. The number of Active Learning stages and the threshold of accepting a prediction as a reliable one are the problem's hyperparameters, depending on various factors, including the audio samples count, the data separability, the classifier's accuracy, and the classification's certainty. The problem's hyperparameters got adjusted using a validation set, $10\%$ of the unlabeled samples. The major innovation in our method is that it overlooks labeling outlier samples until the last stage of labeling because we believe the presence of outlier data samples among the training samples decreases the labeling accuracy. In the first stage of Active Learning-based labeling, we train a classifier using human-labeled data. Then the trained classifier predicts the labels of dataset samples. Next, the reliable predictions reach the final prediction list and the set of training samples for the classifiers of the further labeling stages. In the last stage of the labeling procedure, the labels predicted for the outlier samples and the remainder of unlabeled data reach the final prediction list unconditionally.

\subsection{Data Classification}
\begin{figure}
	\centering	
	\includegraphics[width=\linewidth]{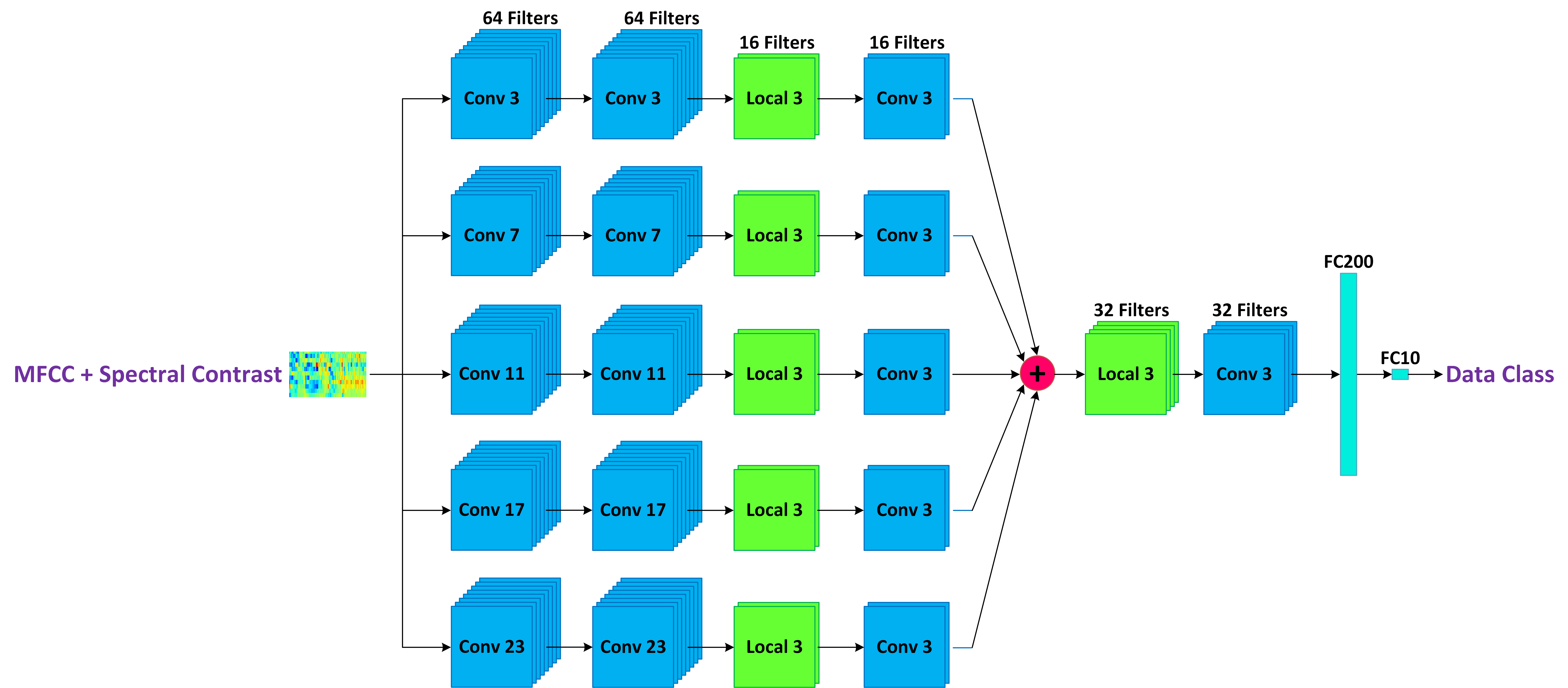}  		
	\caption{The proposed Neural Network for the audio classification task}
	\label{fig: nn}
\end{figure}

Figure \ref{fig: nn} shows the proposed neural network for audio classification. Figure \ref{fig: nn} demonstrates the convolutional blocks in blue and the locally connected convolutional blocks in green. The locally connected convolutional blocks work similarly to convolutional blocks; however, unlike convolutional blocks, a different set of filters is applied at each input patch selected by the kernel window. The number on each block indicates the kernel size for that block, and the number above each column of blocks shows the number of filters. All blocks enjoy Batch Normalization, followed by a LeakyReLU activation function. We merge the outputs of five parallel streams, depicted in Figure \ref{fig: nn}, by an addition operation. The last layers of the neural network are two fully-connected layers, joined together through a dropout layer. We adjusted the neural network's hyperparameters, including filter count, kernel sizes, dropout ratio, and the number of training epochs, using a validation set containing $10\%$ of training samples. In the neural network design, we preferred in-width extension to in-depth expansion, which makes the neural network operate faster by providing the opportunity for parallel computations of five parallel streams.

\section{Implementation Results}
\label{results}
Following the convention of this paper's references, \citep{Shuyang, Takagi, Dilek, Malte, Qin, Ji}, we have tested our proposed context-aware annotation method on the UrbanSound8K dataset. As mentioned in Section \ref{method}, we assume a labeling budget of $15\%$, meaning that the labels are deterministic for only $15\%$ of the data before the labeling process. Table \ref{tab: label} shows the labeling accuracy results, indicating $90\%$ of accuracy, which is improved by $29\%$ compared to the best previously reported works. 

\begin{table}
	\caption{Annotation Accuracy Results with a Labeling Budget of 15\%}
	\centering
	\begin{tabular}{lll}
		\toprule
		Method     & Approach     & Accuracy	\\
		\midrule
		SSCDE\citep{Takagi}	& Semi-supervised Learning	& 35	\\
		CRTAL+SSL\citep{Dilek, Malte}	& Active Learning and Semi-supervised Learning	& 50	\\	
		LDAL\citep{Qin}	& Dictionary-based Active Learning	& 55	\\
		DBAL\citep{Ji}	& Dictionary-based Active Learning	& 57	\\
		MAL\citep{Shuyang}	& Active Learning	& 61	\\
		\begin{bfseries}This work (FACE) \end{bfseries}	& \begin{bfseries}Context-aware Active Learning\end{bfseries}	& \begin{bfseries}90\end{bfseries}	\\
		\bottomrule
	\end{tabular}
	\label{tab: label}
\end{table}

\begin{figure}	
	\centering	
	\includegraphics[width=.5\linewidth]{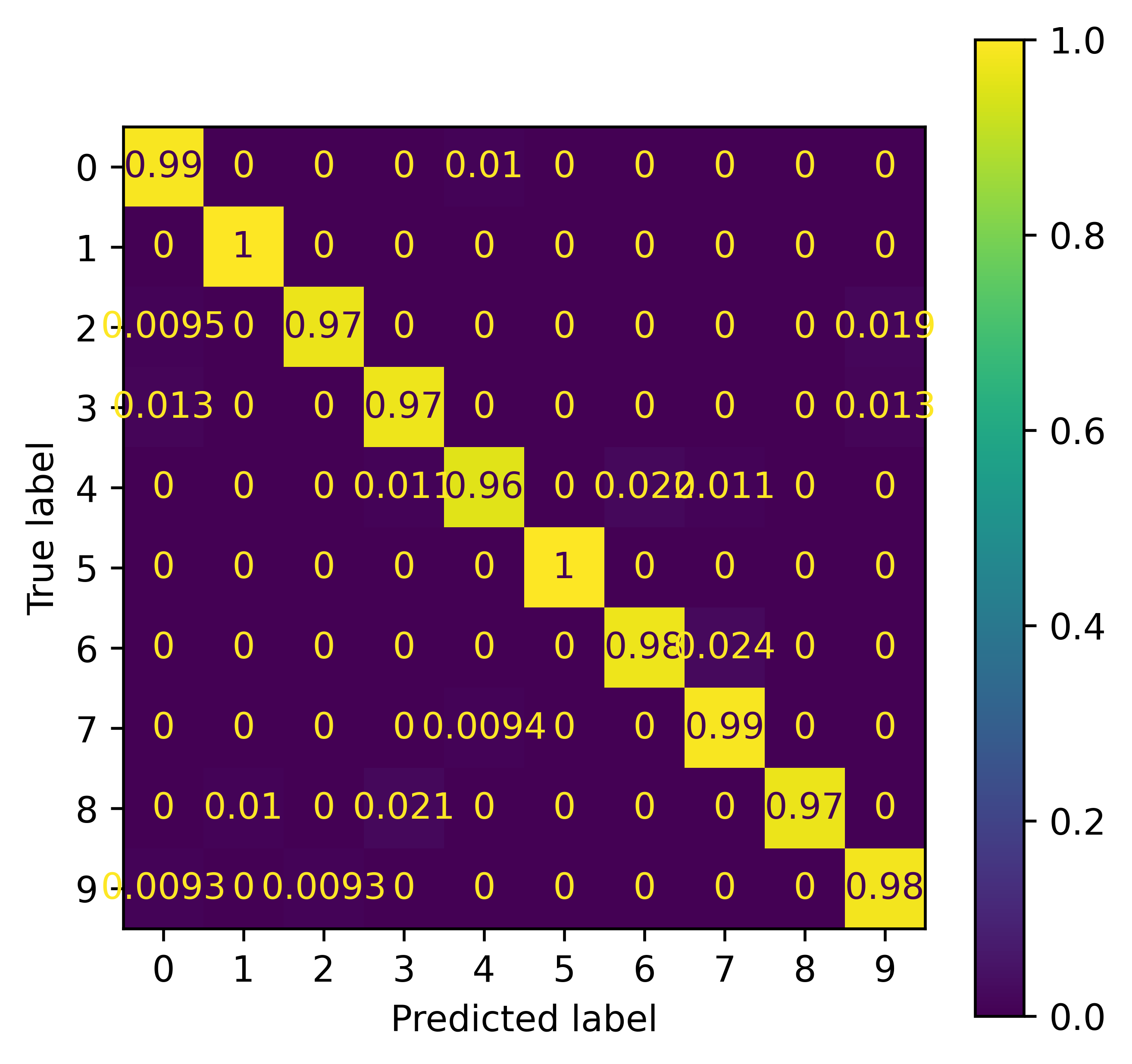}  
	\caption{The classification confusion matrix on the US8K dataset}
	\label{fig: cm}
\end{figure}

For the sound classification task, we trained our neural network in $25$ epochs using a $SGD$ optimizer with a learning rate of $0.01$. We determine the maximum number of training epochs and the optimizer settings such that the classification accuracy on the validation data, $10\%$ of the training set, reaches its highest value. Regarding the popularity of the UrbanSound8K dataset for the audio classification task, Table \ref{tab: class} shows the classification accuracy results on the UrabanSound8K dataset. As the results of Table \ref{tab: class} indicate, our proposed method achieved the classification accuracy of $98.05\%$, which is higher than any other reported works to date. Presenting another measure for our classification method's quality assessment, Figure \ref{fig: cm} depicts the classification confusion matrix for the UrbanSound8K dataset. To study the effect of Active Learning on classification accuracy enhancement, we let the Active Learner, a Python tool from the modAL package, inquire about the labels of some test set samples. We observed that it is possible to achieve $100\%$ classification accuracy by providing the data labels for only $0.9\%$ of the test data.

The notebooks containing our source codes and implementation results are available at \href{https://github.com/gitmehrdad/FACE}{https://github.com/gitmehrdad/FACE.}

\begin{table}
	\caption{Classification Accuracy Results}
	\centering
	\label{tab: class}
	\begin{tabular}{ll}
		\toprule
		Method      & Accuracy	\\
		\midrule
		MAL\citep{Shuyang}				& 65	\\
		CRTAL+SSL\citep{Dilek, Malte}	& 65	\\	
		DBAL\citep{Ji}					& 67	\\
		LDAL\citep{Qin}					& 68	\\
		CDSS\citep{Lee}					& 78	\\
		AemNet-DW\citep{Lopez}			& 83.6	\\
		Graph wo/ FF Ontology\citep{Sun}& 89.0	\\
		TFCNN\citep{Mu}					& 93.1	\\
		SacNet-8\citep{Li}				& 95.3	\\
		TSCNN\citep{Dong}				& 95.7	\\
		TSCNN-DS\citep{Su}				& 97.2	\\
		MHAT\citep{Song}				& 97.5	\\
		\begin{bfseries}This work (FACE) \end{bfseries} 	& \begin{bfseries}98.05 \end{bfseries} 		\\
		\begin{bfseries}This work (FACE + Active Learning) \end{bfseries} 	& \begin{bfseries}100 \end{bfseries} 		\\
		\bottomrule
	\end{tabular}
\end{table}

\section{Conclusion}
\label{conclusion}
This paper presented a fast, accurate, and context-aware method that can get employed to annotate and classify the audio signals. A context-aware approach utilized in feature selection, classifier design, and annotation algorithm design remarkably improves our methods' accuracy in both annotation and classification tasks. The compact extracted features vector and the computationally efficient classification scheme avail a fast labeler, annotating nearly $8000$ audio samples with $90\%$ of accuracy in less than a minute using a free computing platform of Google Colaboratory. Our proposed method obtained an average classification accuracy of $98.05\%$ on the US8K dataset. The achieved classification accuracy is the best-reported one compared to previous works. An Active Learning procedure boosts the classification accuracy to the value of $100\%$ by asking for the labels of only $0.9\%$ of the test data.

\bibliographystyle{unsrtnat}
\bibliography{references}  

\end{document}